# MAGNETIC MEASUREMENTS AT WARM OF THE FIRST FCC-EE FINAL FOCUS QUADRUPOLE PROTOTYPE

M. Koratzinos[1], MIT, G. Kirby, C. Petrone and M. Liebsch, CERN

*Abstract*

The first FCC-ee final focus quadrupole prototype has been designed, manufactured, assembled and tested at warm. The prototype is a single aperture quadrupole magnet of the CCT type. One edge of the magnet was designed with local multipole cancellation, whereas the other was left with the conventional design. An optimized rotating induction-coil sensor was used. A technique was developed to take into account field distortions due to the environment of the test and distinguish them from magnet effects, demonstrating an excellent field quality for the prototype.

## INTRODUCTION

The FCC project aims to deliver a high-luminosity $e^+e^-$ storage ring with a range of energies from 45 to 182.5 GeV per beam (FCC-ee) [1] [2]. It incorporates a "crab waist" scheme to maximize luminosity which necessitates a crossing angle between the electron and positron beams of ±15 mrad in the horizontal plane. The last of the final focus quadrupoles, QC1L1, sits 2.2 m from the IP where the beam separation is only 66 mm. The solution opted for these final focus quadrupoles is a CCT design with no iron yoke and active cross-talk compensation and edge correction [3].

**Table 1:** FCC-ee Final Focus quadrupole vital specs

| | |
|---|---|
| Technology: | CCT |
| Conductor technology: | NbTi |
| Formers: | AL6082-T6, 430 mm length |
| Aperture: | 40 mm |
| Number of conductors: | 8, individually insulated |
| Length of groove: | 15.34/19.16 m (inner/outer) |
| Groove size: | 2.05 × 4 mm |
| Conductor type: | LHC cable 0.825 mm |
| Inductance: | 6.3 mH |
| Magnetic length: | 315 mm |
| Transfer function: | 0.4333 T/A |

The prototype manufactured is a single-aperture quadrupole of 315 mm magnetic length, 26% of the length of QC1L1 [1] which is 1.2 m long, all other parameters being the same (Table 1). It benefits from work done in similar CCT magnets at CERN [4] [5]. All magnetic design was performed using the *Field* suite of programs [6]. The magnet was manufactured in the CERN main workshops. As there is no cross-talk with a single aperture prototype, the compensation idea can be checked by using an edge correction on one side of the magnet and not on the other. The idea behind the edge correction is this: a CCT magnet has non-zero multipole components at the edges, which exactly integrate to zero when integrating over the whole magnet. However, this magnet will be placed in an area of rapidly changing optics functions, and therefore global compensation is not sufficient. Instead, all multipoles vanish locally at the edge of the magnet using the technique described in [3]. *Figure 1* shows the inner magnet former on the corrected edge.

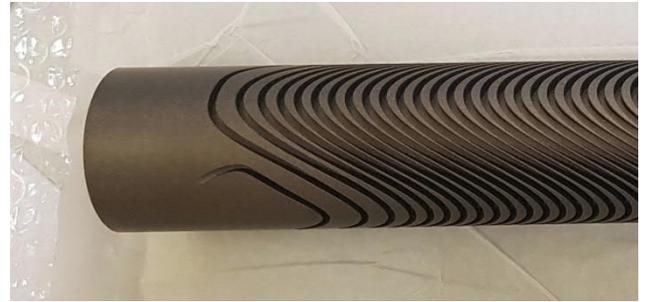

**Figure 1**: *inner former, corrected edge. Note the first two turns that deviate from a pure sinusoidal shape*

The test at room temperature aims to measure the
- Field quality for the bulk of the magnet (excluding the edges), placing the probe at its magnetic centre
- Field quality of the unmodified edge
- Field quality of the modified edge

We can then extrapolate the performance to the full-scale magnet.

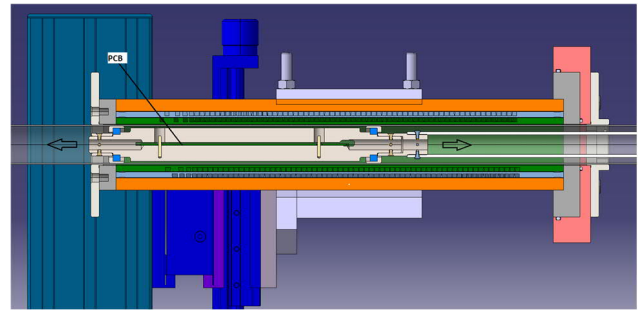

**Figure 2**: *Lateral section view of the designed magnet support and the rotating coil longitudinal scanner system.*

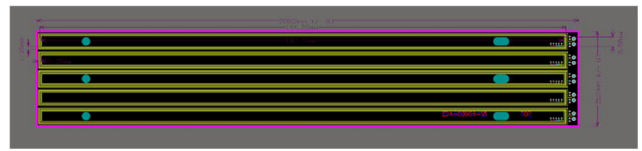

**Figure 3**: *View of the Printed Circuit Board developed for the rotating coil longitudinal scanner system.*

---
[1] Currently at CERN

## TESTING ARRANGEMENT

A dedicated development, profiting from the Printed Circuit Board (PCB) technology, allowed to measure precisely the field using a rotating coil arrangement [7], as shown in *Figure 2*.

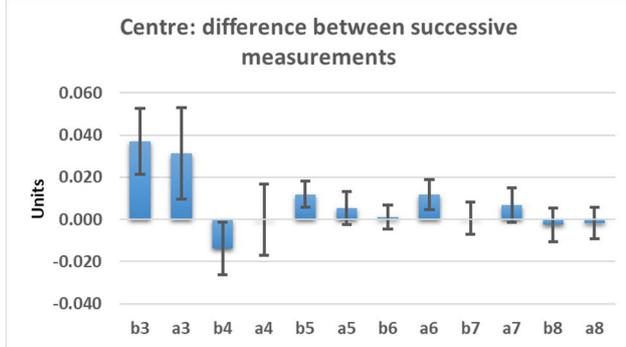

*Figure 4: difference between successive measurements compared to the expected error of the measurement*

*Figure 3* shows the Printed Circuit Board (35 mm width and of active length 194.25 mm) comprising five induction coils. Each measurement averages results over 100 consecutive turns in both magnet current polarities. The measurement precision is shown in the error bars of *Figure 4*. Reversing the current gets rid of any time-independent extraneous field components, like the earth's magnetic field, but it does not get rid of distortions resulting from magnetic objects in the vicinity of the measuring table. The magnet current used is ±5 A, accurately measured and stable at the ~$2 \cdot 10^{-5}$ level, adequate for this analysis. Traditionally, measurements are quoted at 2/3$^{rds}$ aperture: since the beam pipe has a 15 mm radius, all measurement results here refer to a 10 mm as reference radius. The data is post-processed to calculate the 15 multipole coefficients [8]. Results are expressed in *units*: 1 unit = $10^{-4}$ of the main field component.

The first consistency test of the whole chain is the short-term reproducibility. Two measurements were taken 16 minutes apart. Their difference can be seen in *Figure 4*. The error bars are the standard deviations of the distributions of the measurements of each rotation, taken from [7], divided by the square root of the number of measurements, in our case 200. No other systematic errors are added. However, the estimation of error bars seems consistent and the level of agreement of successful measurements is better than 0.04 units.

## MEASUREMENT AT CENTRE

We then measured the multipole components at the centre of the magnet, away from any edge effects.
To make sure that no environmental components affect the measurement, the following approach was adopted: two measurements were taken where the coordinate system of the measurement was not touched, but the magnet and only the magnet was rotated by about 42 degrees. The precise angle is measured by how the pure b2 component of the non-rotated data transformed into a b2 and a2 component in the rotated case. We expect any multipole components coming from the magnet itself to rotate, but any environment distortions to stay as they were, see *Figure 5*. From two measurements, we can derive the two unknowns, the multipole contribution of the environment and that of the magnet. At the level of accuracy of this prototype, this step was essential, as the majority of the measured components in the first measurement did not rotate with the magnet (notably b6, a6) so can be attributed to the environment, possibly due to the presence of high strength stainless steel bolts holding the aluminium frame that supported the magnet in place (see *Figure 2*). Table 2 shows the normal ($b_n$) and skew ($a_n$) multipole errors up to a8 (multipoles up to order 15 were all zero). "Original" and "rotated" are the raw measurements, which were disentangled to the contribution from the magnet (*Figure 6*) and the environment.

**Table 2:** Multipoles (raw measurements and the results of this analysis) in units of $10^{-4}$ for the centre of the magnet.

|    | Raw measurements | | Analysis | |
|----|----------|---------|---------|-------------|
|    | original | rotated | magnet  | environment |
| b3 | -0.220   | 0.083   | **-0.129** | -0.110   |
| a3 | 0.318    | 0.427   | **-0.146** | 0.447    |
| b4 | 0.531    | 0.494   | **0.029**  | 0.510    |
| a4 | 0.538    | 0.656   | **-0.057** | 0.595    |
| b5 | -0.162   | -0.159  | **-0.004** | -0.163   |
| a5 | -0.031   | -0.034  | **0.001**  | -0.035   |
| b6 | 0.642    | 0.650   | **-0.011** | 0.652    |
| a6 | -0.118   | -0.105  | **-0.006** | -0.118   |
| b7 | 0.031    | 0.033   | **-0.003** | 0.033    |
| a7 | -0.004   | -0.005  | **0.001**  | -0.008   |
| b8 | -0.002   | 0.003   | **0.004**  | -0.005   |
| a8 | -0.029   | -0.030  | **0.009**  | -0.036   |

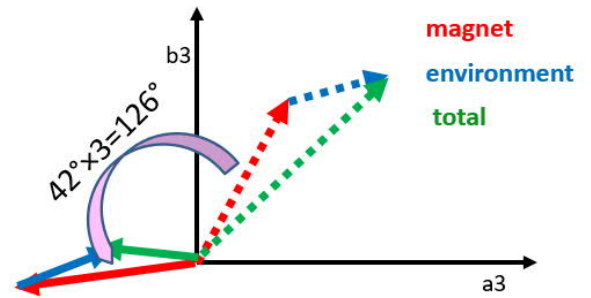

*Figure 5: Explanation of the analysis method used here. In dashed green is the sextupole vector in the a3-b3 plane, comprising the magnet (red) and environment (blue) components. A $42^0$ rotation rotates the magnet component only by $126^0$, resulting in a new total vector (in solid green). The vectors before and after rotation are equal only if the magnet component is zero*

This method cannot be used to extract the two components when the rotation angle times the multipole number is close to 360 degrees. This dictated the choice of 42 degrees that yields good sensitivity.

Note that multipole errors are 0.15 units or smaller, an excellent result, and close to the expected sensitivity of the method.

## EDGES MEASUREMENT

Next task is to measure the multipole components in the two edges, to test the efficiency of the edge correction, only performed on one side. The centre of the probe was placed at ±215 mm from the centre of the magnet. No rotated measurements to discriminate between environment and magnet components have been performed here.

Please note that the multipole components have been normalized to the full magnetic length of QC1L1, the final magnet. The edges quadrupole field integrated over the length of the probe (194 mm) contributes about 1/30th of the field of QC1L1. The comparison of the corrected/uncorrected sides can be seen in *Figure 7*. The correction reduces the effect of the edges by large factors, the resulting multipoles being 0.11 units or less.

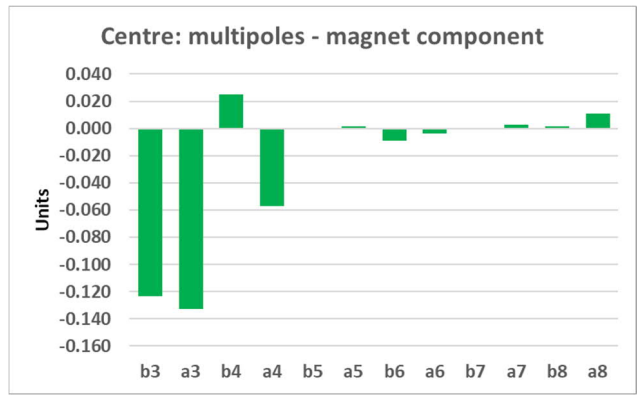

*Figure 6:* multipole components up to a8 in the centre of the magnet, following the analysis described here. All components are 0.15 units or less.

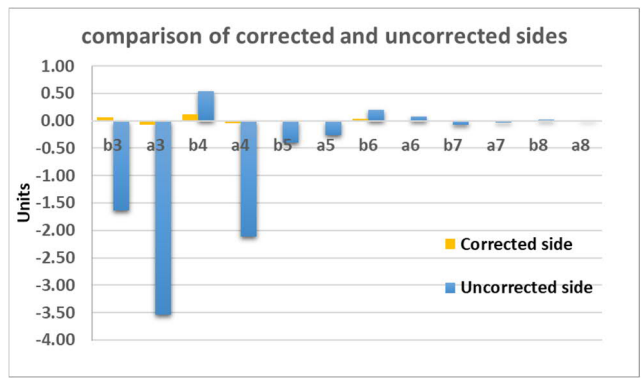

*Figure 7:* comparison of multipole measurements on the corrected and uncorrected edge of the magnet.

## COMPARISON WITH SIMULATION

3D simulation has been performed on the model. For the centre of the magnet, theoretical errors are all zero but for the B6 component, which is expected to be -0.18 units, arising from the finite thickness of the groove. However, there is a competing correction stemming from the fact that the conductor is stiff and slightly smaller than the groove. This produces a positive B6 component compensating the above error (as it turns out, nearly completely in our case).

For the corrected edge, all expected multipoles are zero with the exception of a4 and b6 (due to the fact that the probe is not centred at the magnetic edge of the magnet (-157mm), but its physical edge (-215mm from centre). The only multipoles with some discrepancy are b3, a3 and b4 with differences respectively of (0.06, -0.07 and 0.10) units. This very small difference could be attributed to a possible distortion of the field due to the environment (note that no second measurement at a 42-degree angle was performed here).

The uncorrected edge has not been adequately simulated for a comparison (due to the presence of the splice box).

**Table 3:** measured and expected multipoles on the magnet edge corrected according to **[3]**.

| Multipoles, corrected side (units $10^{-4}$) | | |
|---|---|---|
| order | B components | A components |
| 3 | 0.06 (expected 0.00) | -0.08 (expected 0.00) |
| 4 | 0.11 (expected 0.00) | -0.04 (expected -0.03) |
| 5 | -0.01 (expected 0.00) | 0.00 (expected 0.00) |
| 6 | 0.04 (expected 0.01) | 0.00 (expected 0.00) |
| 7 | 0.00 (expected 0.00) | 0.00 (expected 0.00) |
| 8 | 0.00 (expected 0.00) | 0.00 (expected 0.00) |

## CONCLUSIONS

Tests at warm assessing the field quality of the first FCC-ee final focus quadrupole prototype have been performed. A rotating magnet technique was used to isolate the magnetic errors of the device and eliminate field distortions due to the surroundings.

The quality at the centre of the magnet is 0.15 units or better for all multipoles.

This magnet contains a novel edges correction, where one side of the magnet has been corrected locally. This correction has worked very well, reducing errors due to edge effects by large factors compared to the uncorrected side. The corrected edge has multipole errors of 0.10 units or less.

This demonstrates the suitability of the CCT technique for very accurate accelerator magnets and the correctness of the edge correction technique proposed in [3].

## ACKNOWLEDGEMENTS

We gratefully acknowledge the help of the TE-MSC-ME group of CERN for executing the warm tests and providing us with the raw data. This work would not be possible without the help of Jeroen van Nugteren, Austin Ball, Katsunobu Oide, Frank Zimmermann, Guenther Dissertori, Michael Benedikt, Herman Ten Kate, Karol Scibor and Maf Alidra whose help is gratefully acknowledged.